# Understanding the use of two integration methods on separable first order differential equations

Katrina E. Black[*], Michael C. Wittmann[*†§]
[*] Department of Physics and Astronomy
[†] College of Education and Human Development
[§] Center for Science and Mathematics Education Research
University of Maine, Orono ME 04401 USA
Corresponding author: wittmann@umit.maine.edu

**Abstract:** We present evidence from three student interactions in which two types of common solution methods for solving simple first-order differential equations are used. We describe these using the language of resources, considering epistemic games as particular pathways of solutions along resource graphs containing linked procedural and conceptual resources. Using transcript data, we define several procedural resources, show how they can be organized into two facets of a previously described epistemic game, and produce a resource graph that allows visualization of this portion of the epistemic games. By representing two correct mathematical procedures in terms of shared resources, we help clarify the types of thinking in which students engage when learning to apply mathematical reasoning to physics and illustrate how a "failure to connect" two ideas often hinders students' successful problem solving.

## Introduction

In intermediate and advanced physics classes, a high level of facility in mathematics and mathematical reasoning is required to solve problems, and mathematical insights play a fundamental role in developing conceptual understanding of the physics. In many cases, mathematical statements serve as shorthand for conceptual reasoning about the physics. The solution of differential equations (DEs) and the appropriate application of boundary conditions is one of the most fundamental mathematical techniques applied in the study of physics at all levels beyond introductory. At the University of Maine (UMaine), students are first required to solve differential equations in the physics curriculum while taking a second-year mechanics course. The mathematics department's differential equations course is a co-requisite for the course. Typically, roughly half the students in the intermediate mechanics course have taken the differential equations course, and half are taking it concurrently. In this paper, we describe how students' use of mathematical and physical reasoning tools and skills guide thinking about integration in the context of solving differential equations in intermediate mechanics. Our work underlies the ongoing development of the *Intermediate Mechanics Tutorials* [1,2], though we do not describe the curriculum here. Related work has been reported on previously [3-5].

We are interested in first-order, separable, differential equations (FOSDEs). Unlike many of the other DEs encountered by mechanics students, the solution to a FOSDE is frequently found directly by separation of variables and integration, rather than, for example, the "guess and check" method often used to solve second order DEs. Our students overwhelming prefer to use separation of variables to solve FOSDEs, although rarely they use guessing or the more complicated variation of parameters. Because of this, we limit our discussion to the use of separation of variables. In this case, there are two relatively straightforward methods for applying boundary conditions to the differential equation. First, one can use the boundary condition to find the value of the undetermined constant present in the general solution. (Of course, this is the only method available when the solution to the differential equation is guessed.) Since in the solution to a FOSDE, this constant is the integration constant required by the indefinite integration of the separated DE, we refer to this as the *+C method*. The second common solution method involves using the boundary condition to reason about appropriate integration limits for the separated DE and then using definite integration, bypassing the need to add an integration constant and find its value. We call this the *limits method*.

In our experience, math classes (as is the case at UMaine) typically focus on the +C method, with little or no discussion on the use of limits. While the +C technique is more general mathematically (as a family of solutions is first found, followed by narrowing to a particular solution), physicists often prefer to use limits, relying on physical reasoning to eliminate additional mathematical steps. The limits method also allows for quicker investigation of the dependence of a solution on its boundary conditions, letting limiting cases be tested before a final solution is written. The two methods are procedurally different but, of course, lead to physically and mathematically equivalent solutions. However, as we will discuss, students often do not see the equivalence of the two methods.

Our data come from a variety of interview and group work situations that are described below. We do not emphasize issues of social interaction, attending instead to the connections between math and physics reasoning. Though we are in physics, work by Zandieh [6], Rasmussen [7], and Donovan [8] in mathematics education helps us make sense of student work in applying mathematics in physics.

## Summarizing Resource Theory

Resources are part of a constructivist "Knowledge-in-Pieces" [9,10] approach to modeling learning that has been applied (primarily in physics) to describing conceptual understanding [11,12], problem solving [13], epistemology [14-18], and issues in framing [19]. A resource is an idea or a tool that one uses in problem solving. It has no inherent correctness, though it might be used inappropriately. Typically, we think of resources as some small part of a larger set of ideas. The larger network of resources used in solving a problem can be represented by a resource graph showing resources connected together in a "ball-and-stick" diagram [20]. Using resource graphs, we have represented different forms of conceptual change [21] and explained elements of instruction not made explicit by authors [20]. In previous work, researchers at UMaine have described some heuristics for defining resources and have defined the plasticity of students' use of mathematics in a mechanics class [5,22,23]. In this paper, we extend the resources framework [24] beyond the conceptual and epistemological resources which have previously been described.

Since we cannot observe what students are thinking, we must instead infer what they are thinking as they solve a problem or communicate an idea to someone else. One might look at what students are doing (the steps in a solution, for example) and consider these procedural steps to be procedural resources. Each step is an epistemic action, meaning that it leads to some new information that was not present before the step was carried out. Much like conceptual resources, they are small steps in a larger solution. In this paper, we create resource graphs that consist of procedural resources [25] which represent the epistemic actions students use in solving a FOSDE problem and connect these to the conceptual resources they bring to two different and equally valid solution methods. The set of procedures used to solve a problem can be thought of as a script or an epistemic game [26-30] which, when run completely, accurately, and appropriately, leads one to the solution of a problem.

In the next section, we present transcribed verbal data from students reasoning about air resistance problems. We reserve our analysis of student resource use until later in the paper, first giving examples from a series of interviews and a videotaped group quiz.

## Three group interactions when setting up the integral

In 2006 through 2008 we videotaped students in a variety of tasks, including small group interviews taking place outside of class, tutorial instruction during class, and group quizzes. From these videotapes, we have chosen three instances in which students are using both the +C and the limits method to solve a problem like those given in Figures 1 and 2. These instances are taken from many hours of events, and our description is not meant to deny that other events could have happened at other times.

We present the group interaction data in two parts. In the first section, we describe two different group mini-interviews ("miniviews"). These were 15-20 minute interviews held after lecture on a weekly basis, always with the same group, throughout an entire semester. Three groups participated, one on each day of class, each group with a different interviewer. Different interviewers (including one author, KEB) used a protocol determined by a committee including both authors and two others on a weekly basis. We present data from two groups discussing an air resistance problem.

In the second section, we present transcript from a group quiz in which students were required to solve an air resistance problem similar to the miniview problem. While the miniviews took place toward the beginning of instruction on air resistance, the quiz took place after all instruction on the topic. Because of the testing situation, a different style of dialogue took place – more focused on a solution, we assume.

### Miniviews: Determining limits of integration

The miniview question in Figure 1 was presented to students very early in the semester. The first few steps of the solution were given to the students because we didn't want groups to spend time on the problem set-up in the time-limited miniview; instead, we were focused on how they would go about using the boundary condition once separation of variables had led to two solvable integrals.

> A group is working on the following problem.
>
> A bullet fired horizontally has a muzzle velocity of 366 m/s and experiences a $-cv^2$ air resistance. Find an equation that describes the horizontal velocity of the bullet with respect to time.
>
> A student writes: $\sum F_x = -cv^2 = m\dfrac{dv}{dt}$
>
> $-\dfrac{c}{m}dt = \dfrac{dv}{v^2}$
>
> What would you do next?

Figure 1: A differential equation problem couched as an air resistance problem.

The problem is solvable using either of the two equivalent methods described earlier, the *+C method* and the *limits method*. In both, you carry out separation of variables to arrive at two integrals (as shown in Figure 1, where time and velocity are on separate sides of the equality). In the *+C method*, indefinite integration on each integral is followed by using the initial condition to determine the value of the undetermined integration constant. In this case, the initial condition to be applied is an initial velocity of 366 m/s at time 0 s. In the *limits method*, the same anti-derivatives as in the previous method are used, but the boundary conditions are applied immediately to determine limits of integration. In this case, the time integral runs from time t = 0 s to some unspecified time, *t*, and the velocity integral from 366 m/s to some unspecified velocity, *v*. We use the common physics convention of not using dummy variables, but rather matching the upper limit of integration to the variable present in the differential. It should be noted that *v* represents the velocity at time *t*, i.e., *v(t)*.

Group A (with Derek, Heather, and Wes, all *aliases*[1]) began by successfully solving the problem using the +C method. When they were satisfied with their answer, the interviewer asked if they could solve the problem using limits of integration. The group first (incorrectly) decided on limits of 0 to 366 on the v-integral. Almost as an afterthought, they (correctly) added limits of 0 to *t* on the *t*-integral (for which they used a dummy variable, *s*).

> Wes: So, what are we going to call this? (indicates the upper limit of integration on the *dv* integration) V-one, I mean v-naught? Or 366 period?
> Derek: You tell me.
> Wes: I'm telling you, it's going to be 366. And what's this? (indicates the lower limit)
> Wes: I'm trying to bring the group into this. (He writes zero)

(We ignore several elements of the social dynamic in this and other excerpts, and focus only on the use of integration methods in the transcript.) The students performed the integration, discovering that not only did they have division by zero, but that they were lacking a velocity, *v*, to solve for. They discussed how to change the limits and several suggestions were made.

> Derek: The limits of integration are wrong. You need a variable in there. You have zero to *t* over here, but you have two functions, or two single numbers over there. You need a variable instead.
> Wes: So what do you want to put in?
> Derek: You want to replace zero with a *v*.
> (Later in the interview) Interviewer: Why are you replacing that one with a *v*?

---

[1] Derek and Wes have previously been described in ref. [5].

Derek: Because I'm going to use that as my, uh – Actually, <expletive> I have it backwards. 366 on the bottom and *v* on the top I think.

Derek first attempted to correct both issues with a single change, without thinking about the physical situation. However, a request for explanation caused him to revise his solution to the quandary. Following this excerpt, the group accepted Derek's correct limits (from 366 m/s to *v*) and applied them to the problem.

Group B (with Ben, Kent, Mary, and Ned) began by using the limits method; however, the limits they chose were (the correct) 0 to *t* on the *t*-integration and (the incorrect) 366 to 0 on the *v*-integration, reasoning that the velocity starts at 366 m/s and will eventually become 0 due to air resistance. Unlike Group A, they explicitly let physics guide their use of mathematics.

Kent: I think that, because we have an initial velocity but we do not have a terminal velocity, that we should just do this and then do the boundary conditions as normal.
Mary: The velocity would be negative if we stop at zero.
Kent: Hmm?
Interviewer: What do you mean?
Kent: <reads from problem> Yeah, it's moving faster than – it's going to constantly slow down because there's no force counteracting the force of air resistance, so terminal velocity is therefore zero, assuming that it's going to fall forever.
Ben: So you go from *v*-initial, which is 366, to zero.
Kent: OK, so we have *v*-initial over here and we have a zero. … Umm, and on this side, of course, we have what? Zero to *t*?
Ned: Yep.
Kent: OK

Group B's choice of limits (summarized by Kent) led to the same problems experienced by Group A: division by zero and lack of a *v* variable. Kent's response was to consider a different method.

Kent: Originally, the way I was going to do this was to solve it like a normal differential equation, not do these evaluations at all; set them up as indefinite integrals, but I wanted to see what happened going through it this way.

To Kent, the limits method is a different solution from the "normal" +C method. The group returned to the +C method at this point, which allowed them to reach a successful solution (not shown). Having used the +C method to find a solution, the group revisited their reasoning about the limits:

Kent: At the current time, no matter what the time is, is the velocity automatically zero if it's not initial?
Ben: I don't see what the difference is between what he's trying to do and what we did earlier is.
(later)
Ben: What he's doing is, you're saying is you go from *v*-initial to some *v*. It doesn't go to zero, it goes to some velocity.
Ned: That some velocity is what we care about.

Kent continues to drive the group dialogue, fixing the issue (as Derek had in Group A). Others, like Ben, rethink their answers. Kent considers how to set limits consistently for *v* and *t*. Ben and Ned respond with the correct limits, as well. The transition in Ben's thinking (going from limits "from *v*-initial, which is 366, to zero" to limits "from *v*-initial to some *v*. It doesn't go to zero, it goes to some velocity") illustrates that a new idea was applied in this situation: variables can be used as limits. These same ideas were also observed in qroup quiz situations, such as the one described below.

## Quiz: Debating limits and constants

Although most groups we observed in the miniviews and group quizzes verbally considered whether to use the limits or +C method, these discussions were generally brief and limited to one or two statements asserting the superiority of one method over the other without discussing why this might be the case. In one group, however, an extended discussion took place between two students, each arguing for his preferred method. While we cannot claim that the vocally-explicit reasoning of these two students is also what drives the tacit decisions of others, it does provide one way of looking at why students may choose one method over the other.

In a group quiz setting, students were asked the question shown in Figure 2. They were to find the horizontal velocity with respect to time of a beach ball thrown horizontally, assuming a $v^2$-dependent air resistance force. They were not given an explicit value for the initial velocity, nor was it labeled $v_0$; we were interested in how, if at all, the descriptive phrase "throw it hard" would be translated mathematically. All groups were successful in defining the problem, setting up the basic equations, and working toward a complete solution.

> You are at the top of a cliff and have a beach ball of a size such that only the quadratic (and not the linear) velocity air resistance term exists, $F_{air\ on\ ball} = -bv^2$. You take the ball and throw it hard in the horizontal direction. Assume that during this time it does not fall downward at all, but only moves horizontally. Find the equation for the horizontal velocity of the ball as a function of time.

Figure 2: An open-ended differential equation problem couched as an air resistance problem.

In one group, Group C, two students, Max and Phil (*aliases*, where Max is also described in ref. [3]) set up the equation appropriately and arrived at the separated differential equation shown in Figure 1. (Note that the constant $c$ in Figure 1 was called $b$ on the group quiz.) Next, they indefinitely integrated the $dv$ side without adding a constant of integration or applying limits. Because there was no disagreement and no discussion of this solution, we are unable to comment further on this solution. We begin the transcript at the point where Phil asserts that the next step in the solution is to indefinitely integrate the $t$ equation and add an integration constant:

> Phil: so we get negative $m$ over $b$ times negative $v$ to the negative one equals $t$ plus $c$.
> Max: No, this is where you should be going just plain $t$, or $t$-naught to $t$, or zero to $t$, because initial time is when you throw it to some time, so just go to $t$ instead of $t$ plus $c$. <pause> Because then your integration of time would probably be going from zero, when you throw it, to $t$, some time later. So you don't need $c$ there.

Max clearly objected to Phil's approach. His use of language ("you should be going" or "just go to t") implies that he would prefer using limits of integration, although earlier (transcript not given) he had not objected that limits were not used when performing the $v$-integral. Phil, however, is not convinced.

> Phil: Hmmmmm… or you could do $t$ plus $t$-naught. You need a constant though.

Although Phil offered to change the name of the constant (from $c$ to $t_0$) to indicate that it represents a time, he was unwilling to assign that constant any particular value. Additionally, Phil is still following the formalism of the $+C$ method, adding $t_0$ rather than subtracting it, which would indicate a mental switch to the *limits* method. Max asks:

> Max: Why would you need a constant in time? You just go from 0 to $t$.
> Phil: Because it's <sigh> --
> Max: -- $t$-naught to $t$
> Phil: -- that's not a function, well
> Max: well, you could go from $t$-naught to $t$, but $t$-naught is just initially zero when you start throwing the ball.
> Phil: If you, ssssss. No, you can't do that! Because then you're not setting up a function. You're setting up a function that's only OK in certain cases, you see what I mean?

For Phil, without a constant, you don't have a "function". We infer that what Phil is actually referring to is a family of functions. A function "that's only OK in certain cases" is not general enough. It seems that, to Phil, the integration constant must remain as general as possible. In contrast, Max is trying to use information specific to this physics problem in working through the mathematics.

Later, Phil and Max are still discussing the limits versus integration constant issue. Note that the problem statement does not specifically say that you throw the ball at $t = 0$, which causes problems in the following debate:

> Phil: I just can't imagine integrating without having an extra constant. I guess if you said $t$-naught is zero, so, I mean, yeah
> Max: but $t$-naught is the initial time
> Phil: But you don't know that's zero.
> Max: You can just mark it as zero anyway. <pause> For consistency. Why, why do you need some extra time there?
> Phil: Yeah, but then you're not creating an objective function that anyone can use. You're making a function that *you* know, and that's great, but it's not really, I mean –
> Max: Well, no.
> Phil: – the math has to cover everything. We'll compromise, we'll say $t$ minus $t$-naught. Write that. That work?
> Max: That's fine.
> Phil: OK.

Now, Phil does not object to using limits if they are arbitrary (for example, from $t_0$ to $t$) but still does not want to apply a specific value to the constant. He seems uncomfortable using the common physics convention of calling the initial time zero. Again, he tries to be as general as possible, arguing that the final equation must be one that "anybody can use" (as in, a family of functions) and not "one that *you* know" (in this case, an equation with a specific boundary condition applied). Max, though, is looking for consistency between the physics and the math. As we describe below, their compromise in this last few lines satisfies both their agendas in solving this problem, and they are able to move forward without actually resolving the situation. Unfortunately, since their compromise does not correctly incorporate the initial conditions (specifically, they do not apply similar reasoning to the $dv$ integral), their final solution does not describe the given situation.

**Interplay of Math and Physics Reasoning**

Before we model student reasoning using a resources framework, we summarize some of our observations of the groups in these three examples. Though Groups A and B talked about a velocity integral and Group C talked about a time integral, we can contrast their use of mathematics, specifically how they set up and solved the integrals.

<u>Successful solutions</u>

Although Groups A and B took slightly different paths in their quest for a solution, several similarities can be noted. First, both solved the problem correctly using the +C method. While this may seem glaringly obvious, it is useful to know that the mathematics involved in the solution was not too difficult when familiar solution techniques were used. At this early point in the semester, both groups consider the +C method to be the "normal" method for solving a first-order differential equation (as Kent states). We infer a similar attitude from Group A's actions, since they did not attempt or consider using limits until prompted by the interviewer.

Phil, in Group C, was similarly invested in the +C method. Note that Groups A and B used their +C solutions to aid in their reasoning about integration limits, while Phil used Max's reasoning to modify his use of the constant, instead.

<u>Problems with unexpected limits</u>

As we suspected they might, Groups A and B had initial difficulty reasoning about correct limits. Group A chose what we have found to be the commonly expected limits in a physics problem, that is, they automatically chose 0 for the lower limits of both $v$ and $t$ integrals and picked upper limits from the problem statement: $t$ for the $t$ integral, and 366 for the $v$ integral, since that was the velocity given. We note that the initial, non-zero value of velocity in the problem was chosen on purpose to elicit information about students' responses in unexpected situations.

In contrast, Max, in Group C, was quite careful in his choice of limits, but it took some time for him to arrive at his final response and state it clearly. Additionally, in the integral he discussed, zero was an appropriate lower limit; we cannot know how he would respond in a situation that calls for a non-zero lower limit.

Leaving out variables

Surprisingly, students in both Groups A and B initially left out the velocity variable in their integration limits, using only numerical values in their first attempts. The method of putting the integration variable into the problem as a "running" limit of integration seems unfamiliar for students. Group A used a dummy variable, $s$, on the time integral to avoid this issue, but did not use a dummy variable in the velocity integral. It is possible that our students have never or only rarely considered having a variable as a limit of integration, being more familiar with constants or values as limits.

Reasoning across the equal sign

Students must consider the relationship between the limits on both sides of the equation rather than dealing first with one side and then the other. Because $v$ is a function of $t$, the limits on the $v$ integral cannot be *any* value of $v$; they must be the value of $v$ at the time of the corresponding $t$ limit. Group A did not set limits according to this reasoning, using rules of thumb instead. Time $t = 0$ was matched with velocity $v = 0$, and the variable upper limit $t$ was matched with the constant 366 m/s. Neither of these pairs of limits matched the description of the problem. Also, the dummy variable in the time expression, $s$, was not matched by a similar dummy variable in the velocity expression.

Group B had slightly more sophisticated initial reasoning. Their choice of $v$ limits, from 366 m/s to 0 m/s, does reflect the physical situation described in the problem. However, they still considered the $t$ limits and $v$ limits independently, choosing correct though unmatching $t$ limits of 0 to $t$. It was not until the students considered the functional relationship between each pair of upper and lower limits that they were able to choose those limits appropriately.

Max, in Group C, was very clear about his choice of limits for the time integration, but Group C did not use limits when solving for the $v$ integral. As a result, it is not possible to describe whether Max was taking into account the functional relationship between limits on either side of the equal sign. However, since Max did not reconsider the indefinite integration of the v-integral, it is unlikely that this relationship drove his thinking about the problem.

## Conceptual and Procedural Resources

To model the results shown in the previous section, we use a resources framework that describes both conceptual and procedural elements of student reasoning. We note that most resources discussed in the literature are conceptual or epistemological, while the definition of resources as pieces of knowledge also leaves room for procedural, factual, and analytical resources. We focus on the procedural elements as a way of mapping our steps in an epistemic game [26-29] and illustrate how the procedural resources require certain conceptual resources to be used appropriately. In the discussion below, we describe Max and Phil, adding when necessary a description of miniview Groups A and B to support our interpretation.

### Describing epistemic games as made up of procedural resources

Much as resources are a kind of schema, epistemic games are one kind of script (the former describes conceptually-linked ideas, the latter time-ordered actions). Epistemic games are sets of rules that are followed when creating new knowledge. They have starting conditions, moves, and an ending condition, known as the epistemic form. In the language of the resources model, a student takes a particular path through the activated resources of his or her frame [19,20]. If the links between these resources are robust and often activated in the same order in varying situations, we can call this path an epistemic game. Because the resources activated in a particular situation depend on the setting, the individual, and on other resources already activated, the epistemic games available for play are personally and contextually dependent; more than one game may be available in a particular situation. Similarly, since resources can be applied in a variety of situations, the same game may be available in different situations.

In physics, epistemic games have been described at a fairly large grain size [29]. We seek a smaller grain analysis. In our description, an epistemic game is personal and dependent on the particular resources that are activated. Since the moves in large-scale epistemic games are applicable in widely varying situations, one can average over students rather than time to establish the validity of these games. As the grain size gets smaller, so does the likelihood of any two students having identical resources, and thus, identical games. A complete identification of a small-scale epistemic game must then involve several observations of the same student playing the same game. This is often impractical; however, we can speculate about possible games based on a single observation.

In Group C, both Max and Phil can be seen as playing variations of the large-scale game "Mapping Meaning to Mathematics" [29]. This game has the moves "Develop story" about physical situation, "Translate quantities" in physical story to mathematical entities, "Relate mathematical identities" in accordance with physical story, "Manipulate symbols," and "Evaluate story." Specific details differ for Max and Phil, though, as we describe below. In the miniviews, some moves had already been carried out in the problem statement, so while Groups A and B seem to be playing this game, we cannot know what they would have done without the two lines of mathematical prompting.

To describe both Max and Phil's epistemic games, we build a resource graph out of procedural resources. For example, two procedural resources that students use when integrating a physically meaningful equation are "Find Value" (of an integration constant) and "Choose Limits". How can these be considered resources? They have a binary activation state; they are typically implemented as a whole; and they have no inherent correctness, but are simply tools we use in a given mathematical setting. Each connects the physical meaning of the equation to the mathematical formalism and can involve several sub-steps when applied. These resources are frequently used in conjunction with the procedural resource "Extract Boundary Condition", in which the student uses the problem statement to develop the mathematical form of the boundary conditions.

We make two formal points. First, different procedures may use the same conceptual resources. For example, both "Find Value" and "Choose Limits" might require the use of the symbolic form "Identity" when assigning "Value" to different variables [31,32]. Second, consistent with our general definition of resources, we note that the term "resources" describes constructs and procedures that often require and contain sub-steps that we do not make explicit. For example, the actual process of "Find Value" can be quite difficult algebraically. Obviously, analyses at different grain sizes (e.g., smaller or larger procedural steps) are possible.

**Facets of the Integration Game**

Although Phil and Max have identical end conditions at a large grain size (an equation describing the speed of the ball), in a finer grain analysis, they differ. Phil requires an equation that can "cover everything," while Max questions the need for an unspecified constant. Thus, Phil and Max can be described as playing different versions of one epistemic game, as they are "Making Meaning of Mathematics" differently.

Much as there are different observable facets of a single resource (for an example, see ref. [33]), we observe that there are different facets of epistemic games. We call these versions "Finding a Family of Functions" (the +C method without finding the value of the constant) and "Fitting the Physical Situation" (the limits method). The two have similar moves and are facets of the "Making Meaning of Mathematics" epistemic game. We refer to them as facets in recognition of Minstrell's work, where his facets of reasoning are often (though not always) examples of resources or p-prims in a specific context [34,35].

An example resource graph is given in Figure 3, illustrating the two facets of the "Mapping Meaning to Mathematics" game in the context of integration. These correspond to *Fitting the Physical Situation*, represented by solid arrowheads, and *Finding a Family of Functions*, represented by open arrowheads. For reasons of space, we show only the procedural and not the conceptual resources in this graph; conceptual resources are discussed in the text. Arrows indicate the order in which the resources appear in the game, and the absence of links between resources indicates only that the unlinked resources were not observed as being connected in this particular context, not that no link exists in any context. Finally, we note that plotting other students' methods on this resource graph would likely result in some re-ordering of arrows; however, most students we observe using separation of variables fall into one of the two facets we describe.

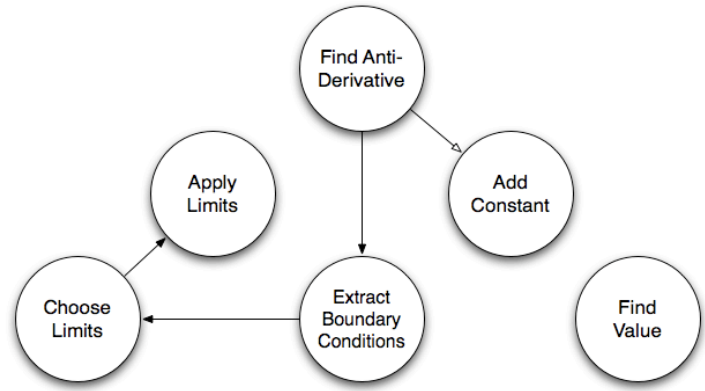

Figure 3: A partial resource graph of two epistemic games. *Fitting the Physical Situation* is represented by solid arrowheads; *Finding a Family of Functions* by open arrowheads.

"Finding a Family of Functions"

For Phil, the game "Finding a Family of Functions" contains only a few procedural mathematical resources, including "Compute Anti-Derivative" and "Add Constant". (See resources connected by open arrows in Figure 3.) He needs only "Add Constant" to make a family of functions from the antiderivative. It is clear that for Phil these resources are very strongly linked. He states that he "just can't imagine integrating without having an extra constant." It is possible to describe Phil's procedural resource "Add Constant" as linked to such conceptual resources as the symbolic form *Base ± Change*. [31,32] In this case, the *Base* is the anti-derivative, while the *Change* is the integration constant that adjusts the anti-derivative to fit a specific boundary condition.

We do not doubt that Phil can access the resources "Extract Boundary Condition" and "Find Value" in other settings, but they are not activated here. He states that Max's suggested boundary conditions cannot be used because "you don't know that's zero." Based on this statement, we suggest that if explicit initial conditions (such as "the ball is thrown with initial velocity $v_0$") were given in the problem statement, his framing of the problem would be such that "Extract Boundary Condition" and "Find Value" would be activated. In contrast, the students in miniview Groups A and B use both "Add Constant", and, later "Extract Boundary Condition" and "Find Value." Such a discrepancy in student methods is one reason for assuming that these are individual procedural resources that must be activated separately and drawn separately in a resource graph.

"Fitting the Physical Situation"

For Max, the game "Fitting the Physical Situation" (see the closed arrowheads in Figure 3) contains a different series of moves, some shared with the previous "Finding a Family" game, but some different. We find evidence for the activation of four procedures.

He joins Phil in "Computing the Antiderivative." In contrast to Phil, Max believes it is necessary to "Extract Boundary Conditions" from the physical scenario and seems guided by a desire for "consistency" (seemingly between the mathematics and the physics).

He makes explicit use of physics conventions, activating conceptual resources such as "Identity" when "Choosing Limits." Although the problem statement does not explicitly state that the initial time is equal to zero, Max feels that an initial time must be defined, and that zero is a conventional initial time for this type of problem.

Max gives several possibilities for limits; it is clear that he needs "Extract Boundary Conditions" to narrow down the options. Eventually, he uses just one set of limits, zero to t, and describes the reason for his choice.

While it is clear that in the context of *Finding a Family of Functions* game, the resources of "Extract Boundary Condition", "Add Constant", and "Find Value" are distinct, their counterparts in the *Fitting the Physical Situation* game, "Extract Boundary Condition", "Choose Limits," and "Apply Limits" may seem nearly identical. However, observations indicate that they are not the same. We have described Max first "Extracting Boundary Conditions" by reasoning about the written physical situation and translating that description into the mathematical description of initial time equaling zero seconds, and then using that information to "Choose Limits" of zero to t. In contrast, students in miniview Group A do not "Extract Boundary Conditions" but rather use surface features of the problem statement combined with a habitual lower limit of zero to "Choose Limits." Group B does "Extract

Boundary Conditions", both for the initial time and as time goes to infinity, but does not have a good understanding how to use these boundary conditions to "Choose Limits".

Although we use Max's actions during the solution of the time integral to describe a possible epistemic game, we have evidence that this game is not necessarily as fully formed as we might hope. Max does not insist on putting limits on the *v*-integral, in fact, he does not even bring the matter up. It is not until Phil wants to "Add Constant" that Max brings up that he may want to "Apply Limits." Thus, we can posit that the resources "Apply Limits," "Choose Limits," and "Extract Boundary Condition" were activated in Max by Phil's use of "Add Constant." Once these resources were activated and all the moves of the limits game made available to Max, he was confident that the limits game was the one he wanted to play.

## Summary

We have extended the use of resource graphs to include procedural resources. We used these resources to describe two facets of a previously documented epistemic game. Our context was the solution of differential equations related to the motion of objects experiencing air resistance. The two solution methods (both potentially correct) share some conceptual and procedural resources. To accurately portray student reasoning, we have looked at three sets of students engaged in dialogue about how to set up (and solve) the equation appropriately. From these examples, we have suggested that the individual steps used to solve the problem can be described as resources, consistent with previous definitions of conceptual and epistemological resources.

We find that students start the semester more familiar with the integration constants method ("finding a family of functions") than they are with the limits method ("fitting to a physical situation"). The former is more common in math classes, while the latter is commonly used in physics and unfamiliar to most students when first encountered. The integration constants method is preferentially employed by two groups at the beginning of instruction on solving air resistance problems, and is still the preferred method for at least one student during a group quiz following instruction in which the limits method was modeled. Students who choose the relatively unfamiliar limits method may have difficulty choosing appropriate limits. Both Groups A and B first chose unphysical limits, failing to consider the functional relationship of the upper and lower limits on either side of the equal sign.

The tools needed for understanding the limits method are only subtly different from those needed to use the integration constants method. In both cases, one needs to extract boundary conditions based on the problem statement. But, we observe differences in student performance that imply that the procedure of "Extracting Boundary Conditions" differs from "Choosing Limits" and "Applying Limits." Evidence suggests that students using limits are thinking of a range (in this case, in time, as expressed with gestures in space, and phrases such as "going from zero to t") while students using the integration constants method are either triggered by mathematical notation or the habits of convention.

## Acknowledgments

We thank Kate McCann, Eleanor Sayre, and Padraic Springuel for their assistance in gathering the video data. This material is supported in part by the National Science Foundation under grants DUE-0441426, DUE-0442388, and REC-0633951.